\documentclass[aps,prb, reprint,
showpacs,superscriptaddress,floatfix,amsmath,amssymb]{revtex4-1}

\usepackage{color}

\newcommand{\sub}[1]{_{\mbox{\scriptsize {#1}}}}

\usepackage{graphicx}% Include figure files
\usepackage{dcolumn}% Align table columns on decimal point
\usepackage{bm}% bold math
\begin{document}

\preprint{AIP/123-QED}

\title[Direct Simulations of Bubble Nucleation]{Direct Simulations of Homogeneous Bubble Nucleation: Agreement with CNT and no Local Hot Spots}% Force line breaks with \\

\author{J\"urg Diemand}
\email{diemand@physik.uzh.ch}%Lines break automatically or can be forced with \\
\homepage{http://www.physik.uzh.ch/~diemand/}
\author{Raymond Ang\'elil}%
\affiliation{Institute for Computational Sciences, University of Zurich, 8057 Z\"urich, Switzerland}

\author{Kyoko K. Tanaka}
\author{Hidekazu Tanaka}
\affiliation{Institute of Low Temperature Science, Hokkaido University, Sapporo 060-0819, Japan}

\date{\today}

\begin{abstract}
We present results from direct, large-scale molecular dynamics (MD) simulations of homogeneous bubble (liquid-to-vapor) nucleation. 
The simulations contain half a billion Lennard-Jones (LJ) atoms and cover up to 56 million time-steps. The unprecedented
size of the simulated volumes allows us to resolve the nucleation and growth of many bubbles per run in simple direct micro-canonical (NVE) simulations while the ambient pressure and temperature remain almost perfectly constant. We find bubble nucleation rates which are lower than in most of the previous, smaller simulations.
It is widely believed that classical nucleation theory (CNT)
generally underestimates bubble nucleation rates by very large factors.
However, our measured rates are within two orders of magnitude of CNT predictions - only at very low temperatures does CNT underestimate the nucleation rate significantly.
Introducing a small, positive Tolman length leads to very good agreement %of the predicted and measured nucleation rates 
at all temperatures, as found in our recent vapor-to-liquid nucleation simulations.
The critical bubbles sizes derived with the nucleation theorem agree well with the CNT predictions at all temperatures. Local hot spots reported in the literature are not seen: Regions where a bubble nucleation events will occur are not above the average temperature, and no correlation of temperature fluctuations with subsequent bubble formation is seen.

\end{abstract}

% PACS, the Physics and Astronomy
                             % Classification Scheme.
%Use showkeys class option if keyword
                              %display desired
\maketitle

\section{Introduction}

Bubble nucleation happens in boiling and cavitation processes in a wide range of contexts and disciplines\cite{Debenedetti1996},
e.g. the electroweak\cite{Dine1992,Megevand2008} and QCD\cite{Kisslinger2005} phase transitions in cosmology, direct dark matter detection experiments\cite{simple,picasso,Pullia2014}, vulcanism\cite{Massol2005} and hydrodynamic cavitation erosion\cite{Brennen2013} and sonochemistry\cite{Suslick1990}. Despite its fundamental importance, the detailed mechanism of bubble nucleation remains unclear\cite{Wang2008} and
accurate predictions of bubble nucleation rates are still not yet possible\cite{Zeng1991,Wang2008}.

In pure liquids, vapor bubbles must form via homogeneous nucleation, which is often suppressed by a large free energy barrier. For this reason it is possible to heat up pure liquids to a superheated, metastable state before stable bubbles form (boiling). Similarly, reducing the pressure below the saturation pressure leads to a metastable, stretched liquid and eventually to its rupture (cavitation)\cite{Debenedetti1996,Brennen2013,Kashchiev2000}.

The most widely used model to predict bubble nucleation rates is the classical nucleation theory (CNT) \cite{Blander1975,Kashchiev2000,Kelton2010,Brennen2013,Kalikmanov2013}. 
More recently density functional theory (DFT)
\cite{Zeng1991,Shen2001}, square gradient theory\cite{Wilhelmsen2014} and some modifications of the classical theory\cite{Delale2003,Shen2003}
have been employed to model the bubble nucleation process.
Measuring bubble nucleation rates in a perfectly homogeneous liquid is very challenging in laboratory experiments\cite{Vinogradov2008}, but 
can be achieved in principle in computer simulations, both with the Monte-Carlo method \cite{Shen1999} and molecular dynamics (MD)
\cite{Kinjo1998,Wu2003,Novak2007,Sekine2008,Wang2008,Tsuda2008,Kuksin2010,Abascal2013,Watanabe2010,Meadley2012,Watanabe2013,Watanabe2014}.
One main conclusion of most of the MD simulations was that CNT generally
underestimates bubble nucleation rates by very large factors.
However, most of the existing simulations use only around 10\,000 or fewer atoms and could be affected 
by their small simulation volumes \cite{Kuksin2005,Watanabe2010,Meadley2012} and
by errors from applying methods like forward flux sampling (FFS) to the bubble nucleation process
(see ref. \cite{Torabi2013} for a discussion of possible problems).
Up to now, most molecular dynamics simulations were only able to accommodate one bubble nucleation event
\footnote{An exception are the recent multi-bubble simulations by Watanabe et al. \cite{Watanabe2013,Watanabe2014}.
They are of similar scale as the ones presented here, however they do probe the spinodal regime and do not measure nucleation rates.},
and with a large number of simulations they can constrain the mean first passage times (MFPT)\cite{Wedekind}. However due to unknown initial lag times
and the early transient nucleation phase, the relation between MFPT and the true steady state nucleation
rate is more complex than usually assumed\cite{Shneidman2014} and steady state nucleation rate estimates based on MFPT
can disagree by several orders of magnitude\cite{Mokshin2014}. Here we present the first direct MD simulations which
are large enough to resolve several bubble nucleation events in the steady state nucleation phase, and allow direct
measurements of the bubble nucleation rates for the first time.

Section \ref{sec:theory} provides a summary of the CNT, section \ref{sec:simulations} describes our MD simulations and analysis methods. In sections \ref{sec:rates}, \ref{sec:crit_sizes}, \ref{sec:size_dist}, \ref{sec:local_hot_spots} we present the results for the nucleation rates, critical sizes, size distributions and our investigation into local hot spots preceding bubble formation. Finally, section \ref{sec:summary} concludes the paper by summarizing our findings. 

\section{Classical nucleation theory}\label{sec:theory}

Classical nucleation theory (CNT) \cite{Blander1975,Kashchiev2000,Kelton2010,Brennen2013,Kalikmanov2013} estimates the work required to form a spherical vapor bubble of radius $r$ under the assumption of mechanical equilibrium,
\begin{equation}\label{DeltaG}
\Delta G (r) =  4 \pi r^2 \gamma - \frac{4 \pi}{3} r^3 (P_{\rm eq} - P_{\rm L}) \delta\; ,
\end{equation}
where $\gamma$ refers to the planar surface tension, and $P_{\rm L}$ is the ambient pressure in the liquid. The vapor pressure in the bubble $P_{\rm V}$ is expected to be slightly smaller than the equilibrium vapor pressure at saturation $P_{\rm eq}$, and the reduced pressure difference is approximated with the Poynting correction factor $\delta$:
\begin{equation}
\Delta P  =  P_{\rm V} - P_{\rm L} \simeq (P_{\rm eq} - P_{\rm L}) \delta\; ,
\end{equation}
and $\delta$ is given by \cite{Blander1975}
\begin{equation}\label{delta}
\delta  \simeq 1 - \left( \frac{\rho_{\rm V}}{\rho_{\rm L}} \right)+ \frac{1}{2}  \left( \frac{\rho_{\rm V}}{\rho_{\rm L}} \right)^2\; ,
\end{equation}
A common assumption is that the pressure in the bubble equals the equilibrium vapor pressure $P_{\rm eq}$, i.e. the Poynting correction is neglected ($\delta = 1$). Far below the critical temperature, the effect of the Poynting correction becomes very small (see Table 1). In this paper we consider both cases: values assuming $\delta = 1$ will be labeled CNT and those with $\delta < 1$ from Eq. \ref{delta} will be called PCNT.

The free energy $\Delta G (r)$ has a maximum at the critical radius $r_c$. 
\begin{equation}\label{rc}
r_c = \frac{2\gamma}{\Delta P} \simeq \frac{2\gamma}{(P_{\rm eq} - P_{\rm L}) \delta }
\end{equation}
and the height of the free energy barrier equals
\begin{equation}
\Delta G (r_c) = \frac{16 \pi}{3} \frac{\gamma^3}{ (\Delta P) ^2}  =  \frac{4 \pi r_c^2 \gamma}{3}  = - \frac{4 \pi/3  \; r_c^3 \Delta P}{2}\; ,
\end{equation}
% temp: used for a presentation slide:
%\begin{equation}
%E_{\rm th} =  \frac{4 \pi r_c^2 \gamma}{3}  + \frac{4 \pi}{3} r_c^3 \rho_{\rm V} h
%\end{equation}
which is simply one third of the surface term from Eq. \ref{DeltaG} at $r_c$ or minus one half of the volume term at $r_c$.

Small bubbles ($r < r_c$) are understood to form as density fluctuations in the liquid, they are short-lived and are found at any time at a number density of
\begin{equation}\label{sizedis}
n(r) = n_{\rm L} \exp \left( - \frac{\Delta G (r) }{ kT} \right) \; ,
\end{equation}
where $n_{\rm L}$ is the number density in the liquid phase.

The steady state homogeneous nucleation rate $J$ is proportional to the abundance of critical bubbles times a kinetic factor\cite{Blander1975,Brennen2013}
\begin{equation}\label{J}
J = n_{\rm L}    \left[ \frac{2\gamma}{\pi m} \right]^{1/2} \exp \left( - \frac{\Delta G (r_c) }{ kT} \right) \; .
\end{equation}
The pre-factors in equations \eqref{sizedis} and \eqref{J} have no rigorous justification and range of other expressions are sometimes used.
The different pre-factors only differ by factors of a few, which is small compared to the large uncertainties in the exponential factor
(see Baidakov and Bobrov\cite{Baidakov2014} for an overview).

\section{Simulation and analysis methods}\label{sec:simulations}

\subsection{Simulation code, setup and parameters}

The simulations were performed with the Large-scale Atomic/Molecular Massively Parallel Simulator (LAMMPS) code\cite{LAMMPS}. We use a truncated force-shifted Lennard-Jones (TSF-LJ) potential 
\begin{equation}\label{tsflj}
u_{\rm TSF}(r)  = u_{\rm LJ}(r) - u_{\rm LJ}(r_{\rm cut}) - (r - r_{\rm cut}) \frac {d u_{\rm LJ}(r_{\rm cut})}{dr} \; , 
\end{equation}
for $r \le r_{\rm cut}$ and $u_{\rm TSF}(r)= 0$ for $r>  r_{\rm cut}$.
$u_{\rm LJ}(r)$ is the widely used 12-6 Lennard-Jones (LJ) potential
\begin{equation}\label{lj}
\frac{u_{\rm LJ}(r)}{4\epsilon} = \left(\frac{\sigma}{r}\right)^{12} - \left(\frac{\sigma}{r}\right)^6 \; .
\end{equation} 
At the cutoff radius $r_{\rm cut} = 2.5 \sigma$, both the truncated force-shifted potential $u_{\rm TSF}$ and its first derivative vanish continuously. This interaction potential is the same as used in two recent studies of bubble nucleation based on smaller simulations with a few thousand atoms \cite{Wang2008,Meadley2012}. Our simulations contain $N = 2^{28} = 536\,870\,912$  particles.

Four sets of simulations were performed, one to probe the boiling regime at superheated conditions at $T = 0.855 \epsilon/k$ (same as in refs. \cite{Wang2008,Meadley2012}), two in the cavitation regime at $T = 0.7 \epsilon/k$ (the same as in ref. \cite{Meadley2012}) and at $T = 0.6 \epsilon/k$, where negative pressures are required to trigger bubble nucleation, and one in between at $T = 0.80 \epsilon/k$.
The critical temperature of the TSF-LJ potential with $r_{\rm cut} = 2.5 \sigma$ is $T_c = 0.935 \epsilon/k$ [\cite{Errington2003,Wang2008}]. For typical simple fluids, the onset of homogenous superheated boiling requires roughly $T > 0.9 \epsilon/k$ [\cite{Brennen2013}]. Argon system units can be defined by matching the critical temperature: for the full standard LJ-potential $\epsilon/k = 119.8 $K, for the TSF-LJ potential used here one gets
$\epsilon/k = 161.3 $K, $\sigma=3.405$\AA, $m= 6.634 \times 10^{-23}$g and $\tau = \sigma\sqrt{m/\epsilon} = 1.86$ ps, when converting to SI units.

\begin{figure}
\includegraphics[height=.30\textheight]{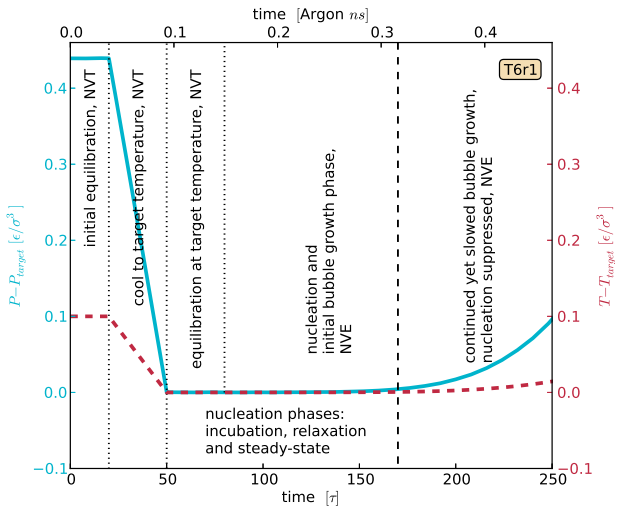}
\caption{(Color online) The temperature (dashed line) and pressure (solid line) differences from the targets for the setup and measurement phase of our shortest simulation, T6r1. In most runs the measurement phase is far longer. In this case, the rapid growth of only a few bubbles quickly leads to a significant pressure increase. Despite this, we are able to measure accurate nucleation rates, shown in the upper panel of figure \ref{fig:treshT6}. Size distributions and formation rates discussed in this paper are taken only during the simulations' `measurement' phase - after the setup, yet before the pressure increase.}
\label{fig:thermo}
\end{figure}

The simulation box is a fixed cubic volume with periodic boundaries. We use the standard velocity-Verlet (also known as leap-frog)
integrator and the time-steps are set to $\Delta t = 0.0025 \tau \simeq 4.65$ fs. 
The liquid was first equilibrated in a stable state at a fixed temperature (e.g. $T = 0.95 \epsilon/k$) for 
10\,000 time-steps. Then the temperature was reduced linearly to the target temperature of (e.g. $T = 0.855 \epsilon/k$)
over 15\,000 time-steps and kept fixed at this temperature for another 10\,000 steps. During
this entire setup phase the average temperature in the entire simulation box was controlled by
simply rescaling all the particle velocities at every time-step. After this setup period the velocity
rescaling was turned off and the runs were continued as micro-canonical simulations, i.e. with a constant number of particles, constant volume and constant total energy (NVE):
We simply and directly integrate the classical equation of motion of the atoms. No artificial constraints like thermostats or barostats are used during our simulations, and we are able to avoid
applying such methods developed for systems near thermodynamic equilibrium to the highly non-equilibrium process of bubble nucleation.
Figure \ref{fig:thermo} illustrates the pressure and temperature evolution over the simulation setup phase.

Energy is conserved very accurately over long timescales: in our longest run (T85r4), the total energy decreased over 1.3 million time-steps, yet only by $1.7\times10^{-7}$ times the initial energy. For our low temperature T6 and T7 runs we use larger time-steps of $\Delta t = 0.004 \tau$. The total energy is still conserved within less than $2.6\times10^{-5}$ times the initial energy in all these runs. While no large, stable bubbles nucleate the thermodynamic states of our systems remain
constant: at the end of run T85r4 the average temperature is $T = 0.85510 \epsilon/k$ and the average pressure increased by only $0.22\%$.

\subsection{Bubble nucleation and growth}

When bubbles nucleate and surpass the critical size, they quickly reach a linear growth regime, where their radius grows at a constant, rather high speed:
 $v \simeq 0.03\sigma / \tau$ at $T = 0.855 \epsilon/k$ and even faster, $v \simeq 0.3 \sigma / \tau$, at $T = 0.7\epsilon/k$.
%Growth, properties and evolution of these bubbles will be analysed in detail in a subsequent paper (Ang\'elil et al. in prep.).
The quickly growing bubbles will at some point start to occupy a significant fraction of the total simulation volume and
the pressure in the simulation box eventually increases. However, our simulation volumes are large enough to allow the formation of many critical size bubbles without a significant pressure increase in most simulations. Due to the relatively large sizes of critical bubbles and the fast growth of bubbles above the critical size, direct NVE simulation of several bubble nucleation events at constant pressure requires very large simulations, and is just becoming feasible with the particle numbers we use here. For the nucleation rate analysis we only use the simulation period, where the total pressure remains within a few percent of the initial
pressure. The tolerance could be set to less than $4\%$ for most runs, but for runs T85r1 and T7r2 it needed to be increased to $10\%$.
to get a long enough period containing several bubble nucleation events. The end of this period is indicated by vertical dashed lines in Figure \ref{fig:thermo}
and Figures \ref{fig:treshT855} to \ref{fig:treshT6} and the average temperatures and pressures during the analysis period are given in table \ref{tab:t3}.

Using a large number of smaller NVE simulations (with up to 11.7 million particles) it is possible to measure the average time required to see {\it one} bubble nucleation event, however one cannot continue into the steady state nucleation regime, which affects the accuracy of the measured bubble nucleation rates\cite{Watanabe2010}. Direct NVE simulations of vapor-to-liquid simulations on the other hand are a bit less demanding, with similar particle numbers as here
($N \simeq 10^9$) one can obtain excellent liquid droplet statistics and very accurate nucleation rate measurements in the steady state regime \cite{paper1,paper2,paper3}.

After the `measurement' phase, when the volume occupied by bubbles becomes larger, the average pressure increases significantly (see Figure \ref{fig:thermo}).
The potential energy becomes more negative as the remaining liquid settles into a slightly denser, energetically more favorable configuration. 
At the same time the average temperature increases and the total energy is conserved very
accurately, even during this very dynamic phase, far from thermodynamic equilibrium. 
% Later, larger bubbles eventually start to oscillate in size, some undergo complete collapse and rebound, and smaller, but initially stable, bubbles disappear completely since the fall below the larger critical sizes
% resulting from the larger pressures.

\subsection{Void identification and bubble properties}

\begin{figure}
\includegraphics[height=.35\textheight]{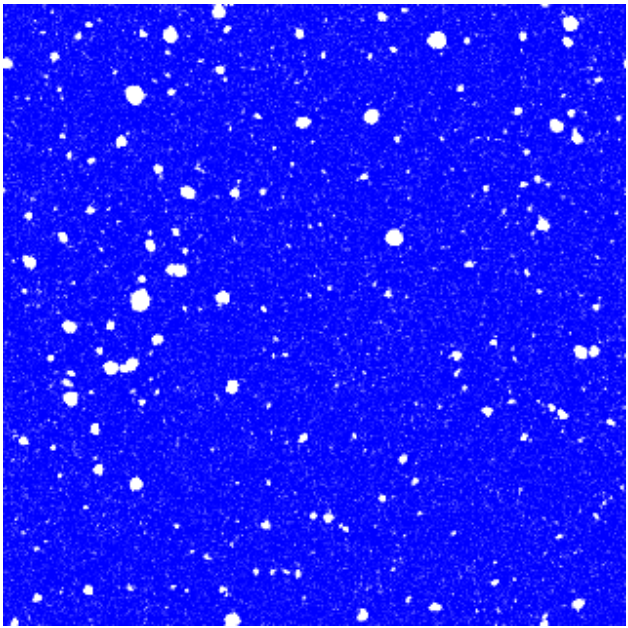}
\caption{(Color online) Projection through run T8r1 at $t=230 \tau$. Every pixel corresponds to a column of 319 cells of
size $(3\sigma)^3$. A larger number of cells below the threshold density of 0.2 $m/\sigma^3$ results in a lighter color (grayscale).
%To reduce the noise in this image eight projections between $t=229 \tau$ and $t=231 \tau$ were averaged.
This run contains about 120 large, stable bubbles at this time, see Figure \ref{fig:treshT8}.
See Supplemental Material for animations of such projections\cite{suppl}.
}
\label{fig:bubbles}
\end{figure}

Vapor bubbles are identified with the following algorithm (adapted from refs.\cite{Watanabe2010,Watanabe2013,Watanabe2014}): The simulation volume is covered with a grid of cubic cells of size $(3 \sigma)^3$. At regular intervals (usually every 1\,000 time-steps) the positions of all cells, which have a number density smaller than $0.2 \sigma^{-3}$ (i.e. less than 6 atoms within $(3 \sigma)^3$) are written out (see Fig. \ref{fig:bubbles}).
Since $\rho_V \ll \rho_L$ at all temperatures simulated here, the identification of vapor bubbles is simple and robust. It works well for a range of threshold densities between $\rho_V$ and $\rho_L$: We confirmed that using other density thresholds ($0.15 \sigma^{-3}$ and $0.3 \sigma^{-3}$) does not
affect the resulting nucleation rates, this was found also in ref. \cite{Watanabe2010}. Using different cell sizes ($(2 \sigma)^3$ and $(4 \sigma)^3$) does not affect the measured rates either.

Then low density cells are linked together into individual bubbles by iteratively checking for nearby low density cells at the 26 neighboring positions in the cubic grid. The resulting bubble volumes are stored and used for the nucleation rate measurements. We also analyzed run T855r2h using smaller cells $(2 \sigma)^3$ and found the same nucleation rate.

For a more detailed analysis of bubble properties some full simulation snapshots are written out. Bubbles near and below the critical size have significantly non-spherical, complicated shapes (as described in ref. \cite{Meadley2012}), while the larger bubbles are roughly spherical. The vapor-liquid transitions are smooth and wide, as found in DFT\cite{Shen2001} and square gradient\cite{Wilhelmsen2014} calculations.
Comparisons with bubble density profiles show that our volume estimates $V$ based on the number of low density cells are reliable: The resulting spherical radii $r_{\rm MD} = (3 V / 4 \pi)^{1/3}$ lie only slightly
below the equimolar radii, which would be the radius of a constant density bubble with a sharp interface (where the density jumps from the central value $\rho_V$ to 
the bulk liquid value $\rho_L$) and the same 
integrated mass\cite{Kashchiev2000}.
% THIS WILL GO INTO PAPER 2:
% The full width of the transition regions turns out to depend only on temperature and to be practically independent of
% superheating (in agreement with \cite{Shen2001}) and bubble size. Bubbles seem to differ
% in this regard from liquid-like clusters, which have size dependent interface widths \ref{paper2}. 
% We find full widths of about $5.0 \sigma$,  $XXX \sigma$ and $2.5 \sigma$ at $kT / \epsilon = 0.855$, 0.8 and 0.7.
% For comparison with the size distributions predicted by the CNT (assumes a sharp interface) 
The differences depend on temperature, but not on bubble size: At $T = 0.855 \epsilon/k$ the equimolar radii are 0.56 $\sigma$ larger than $r_{\rm MD}$, and at
$T = 0.8 \epsilon/k$ the difference is 0.35 $\sigma$. We add these differences to our measurements of $r_{\rm MD}$ when we compare to the size distribution predicted by
classical models in Section \ref{sec:size_dist}. Detailed bubble properties are presented in a subsequent work (Ang\'elil et al. \cite{Angelil2014}). 

\subsection{Nucleation rate measurements}\label{Jmeas}

Our large scale direct nucleation simulations resolve several nucleation events during the steady state nucleation regime at a nearly
constant thermodynamic state. This allows us to determine the nucleation rates $J$ accurately with the Yasuoka-Matsumoto method\cite{yasuoka,paper1} (also referred to as the `threshold method'), where $J$ is simply given by the slope of a linear fit to 
the number of bubbles $N(>V,t)$ larger than some threshold volume $V$.

\begin{figure}
\includegraphics[height=.72\textheight]{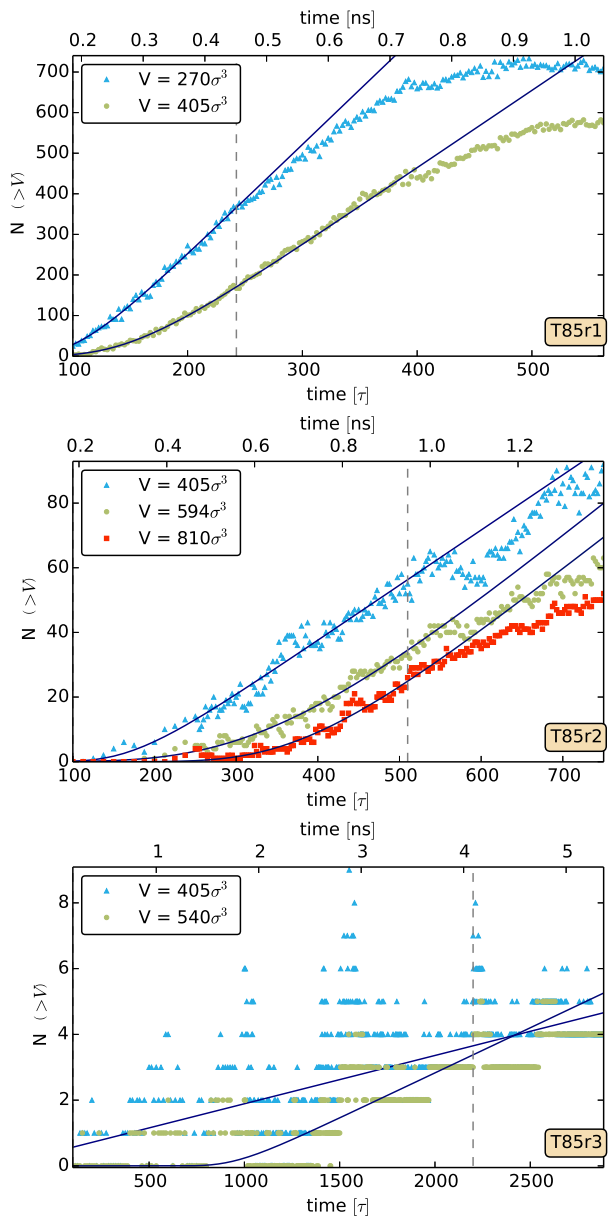}
\caption{(Color online) Number of bubbles above various threshold volumes as a function of time for three runs at $T = 0.855 \epsilon/k$.}\label{fig:treshT855}
\end{figure}

\begin{figure}
\includegraphics[height=.48\textheight]{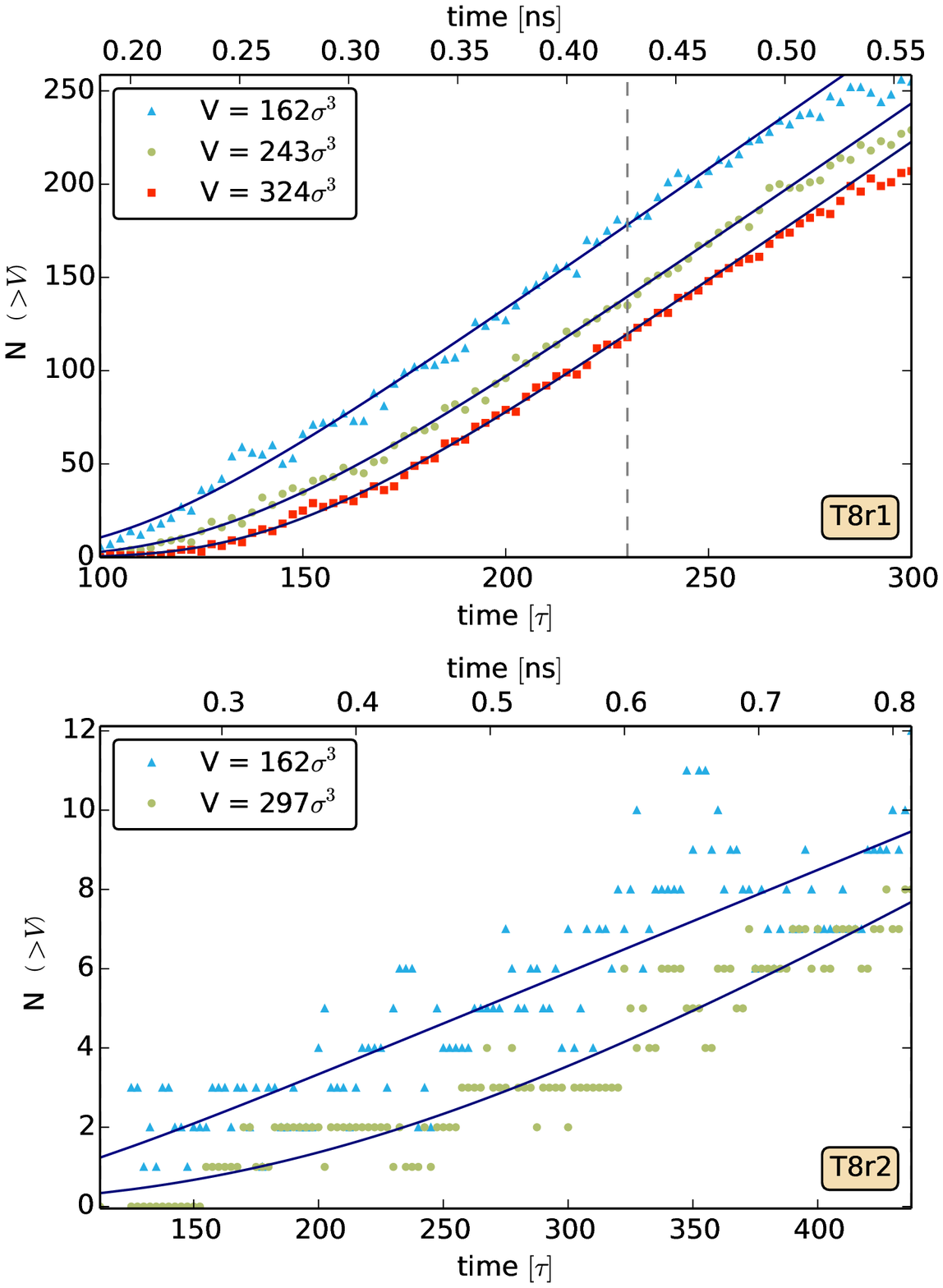}
\caption{(Color online) Number of bubbles above various threshold volumes as a function of time for two runs at $T = 0.8 \epsilon/k$.}\label{fig:treshT8}
\end{figure}

\begin{figure}
\includegraphics[height=.72\textheight]{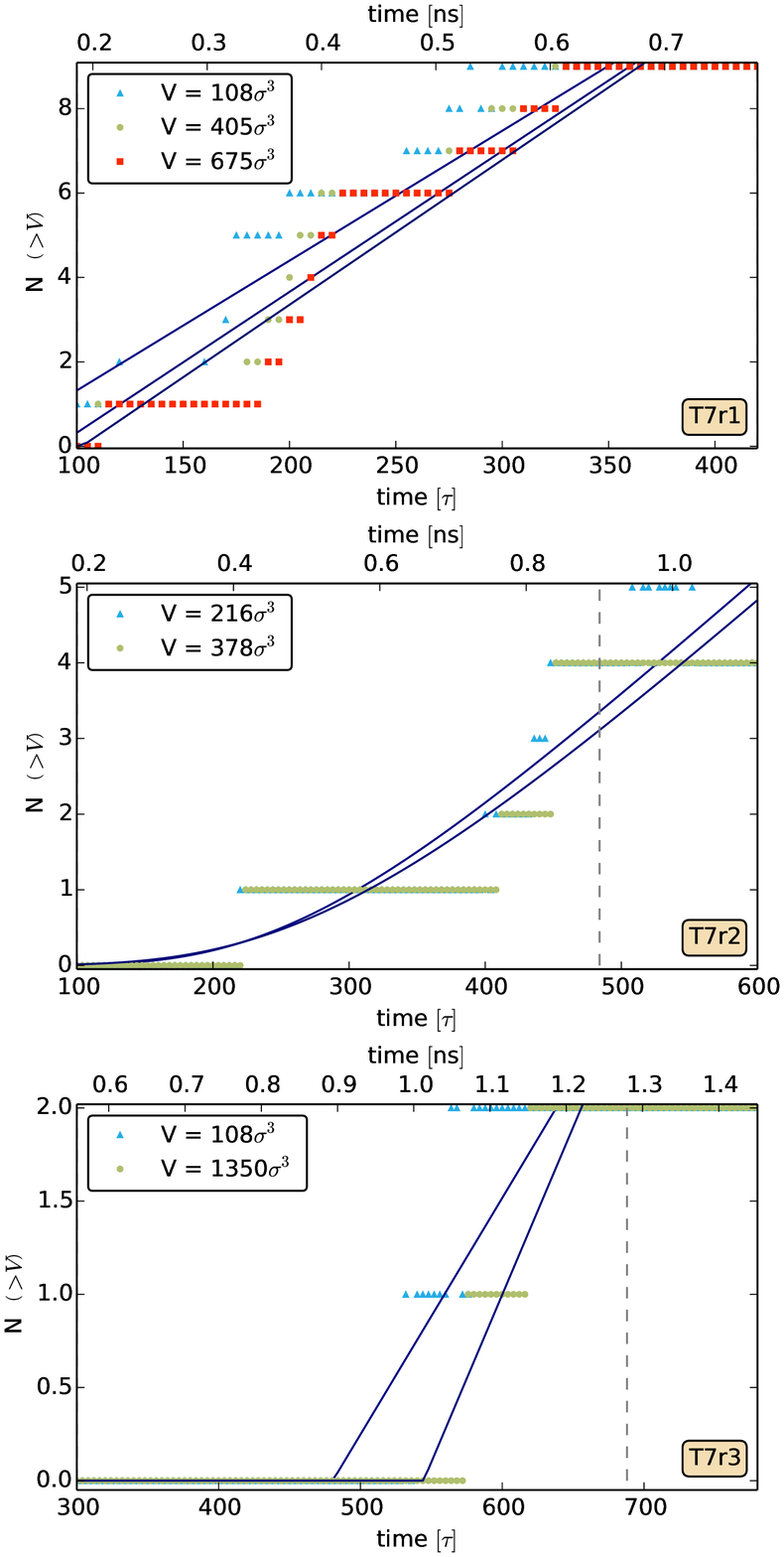}
\caption{(Color online) Number of bubbles above various threshold volumes as a function of time for three runs at $T = 0.7 \epsilon/k$.}\label{fig:treshT7}
\end{figure}

\begin{figure}
\includegraphics[height=.48\textheight]{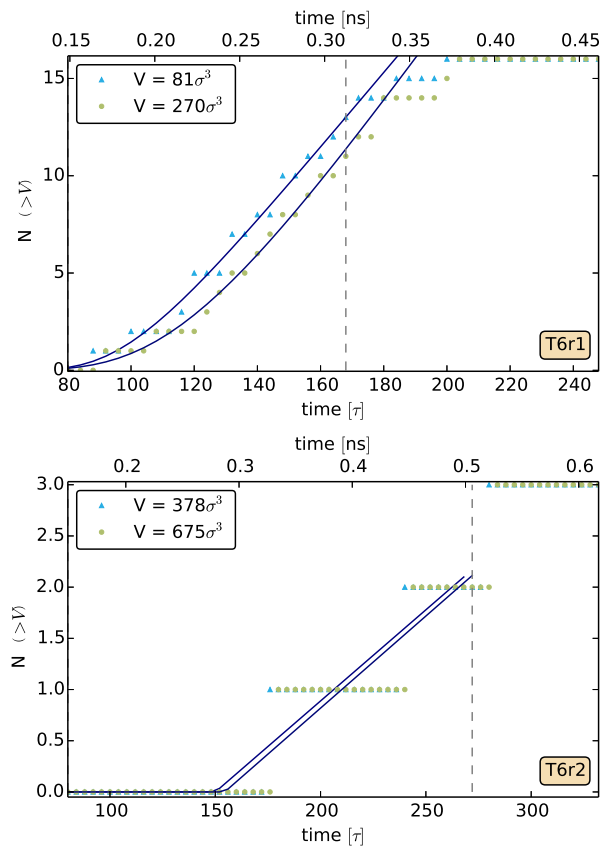}
\caption{(Color online) Number of bubbles above various threshold volumes as a function of time for three runs at $T = 0.6 \epsilon/k$.}\label{fig:treshT6}
\end{figure}

Here we use a non-linear function for $N(>V,t)$ which takes into account the time-dependence of the nucleation rate before the steady-state regime is reached \cite{Shneidman1999,Kashchiev2000,Shneidman2007,Kelton2010}.
It takes some time after the simulation setup until the first bubble nucleation event occurs (called `incubation time' in ref. \cite{Shneidman2007}) and even longer to reach the steady state nucleation regime (related to a `relaxation time', see ref. \cite{Shneidman2007}).
For large times ($t > t_0$) the number of stable bubbles grows linearly: $N(>V,t) = J_{\rm MD} \times (t - t_0) \times L^3$. We find that the function from \cite{Shneidman2007} fits our measured $N(>V,t)$ curves very well, and that the transition timescales (i.e. incubation and relaxation time) are quite long and strongly dependent on liquid temperature and pressure (see Figures \ref{fig:treshT855} to \ref{fig:treshT6}).

We use a least-squares fitting algorithm and minimise the absolute differences in the bubble counts above threshold.
The good statistics of some runs (T8r1 and T85r2 for example), allow a nucleation rate measurement with a relative uncertainty of around $10\%$ and the results are independent of the exact value we choose for the threshold volume $V$. Most runs produce fewer stable bubble and therefore have larger uncertainties in $J_{\rm MD}$ (see table \ref{tab:t3}). Run T8r3 only produced a single stable bubble during the constant pressure epoch, which leads to a large uncertainty in $J_{\rm MD}$, while runs T85r4 and T7r4 did not nucleate at all and only allowed to set upper limits on $J_{\rm MD}$ (see ref. \cite{paper1} for details).

In general, using larger values for the threshold volumes leads to longer lag times before the steady state regime is reached. We therefore use the smaller of the threshold values 
plotted in Figures \ref{fig:treshT855} to \ref{fig:treshT6} to derive the nucleation rate estimates given in table \ref{tab:t3}.

Small simulations are only able to resolve one bubble nucleation event, because the critical bubble volume is comparable to the available simulation volume. With a large number of small runs one can measure the mean first passage time accurately\cite{Wedekind}. But due to the unknown lag time it is not possible to convert mean first passage times into accurate steady state nucleation rates \cite{Watanabe2010,Mokshin2014}. For the case of water vapor-to-liquid nucleation it was recently shown \cite{Mokshin2014}, that the mean first passage time method underestimates the true steady state nucleation rates by about two orders of magnitude (see also ref. \cite{Shneidman2014}). 

\subsection{Convergence tests}\label{sec:conv}

To assess the impact of our chosen simulation box sizes on the measured nucleation rates we performed two additional simulations with the same physical properties as run T85r2, but using
8 and 64 times smaller simulation volumes. Run T85r2h is just large enough to allow a relatively accurate nucleation rate estimate: the total pressure in the box stays within $4.5\%$ of the
initial value until $t = 385 \tau$ and during that epoch we measure and average temperatures and pressures of $T=0.08512\epsilon/k$ and $P=0.01703 \epsilon/\sigma^3$
and we estimate a nucleation rate of $J_{\rm MD} = 2.6\pm0.4\times10^{-10}$.

In the full size run (T85r2) the pressure remains within $3.5\%$ of the target pressure up to $t = 510 \tau$. The larger sample of stable bubbles forming in that period results in a tighter constraint on the
steady state nucleation rate of $J_{\rm MD} = 1.8\pm0.2\times10^{-10}$. The agreement  in $J$ between runs T85r2 and T85r2h within 1.5 times the standard deviation
of the rate from T85r2h is quite good,
especially when the slightly higher average temperature and therefore larger superheating and higher expected nucleation rate in run T85r2h is taken into account.
This shows that our simulations with $N=536\,870\,912$ are large enough and that our nucleation rate estimates have converged. They are not affected by the finite size effects reported by Meadley and Escobedo\cite{Meadley2012} when comparing forward flux sampling (FFS) MD simulations with $N=3\,375$ (the size used in Wang et al. \cite{Wang2008}) and $N=8\,000$. 

Run T85r2q with about 8 million atoms on the other hand is too small to allow a reliable estimate of the steady state nucleation rate. The liquid pressure does increase significantly shortly after the first bubble nucleation event. Reaching the steady state regime and resolving several bubble nucleation events at roughly constant ambient pressure is not possible with direct (NVE) simulations of this size.
 
\begin{table}
\caption{Simulation properties: initial temperature $T$, number of atoms $N$, periodic cube size $L$, atom number density $n$ and total run time.}
\begin{ruledtabular}
\begin{tabular}{ l c c c c c }
Run ID&T&N&L& $n$& $t_{\rm{end}}$    \\
& [$\epsilon/k$] & & [$\sigma$] & $\left[\sigma^{-3}\right]$ & $\left[\tau \right]$
\\
\hline
  T85r1 &0.855& 536\,870\,912 & 987.0&0.558365 & 562.5 \\ %ld4
  T85r2 &0.855& 536\,870\,912 & 984.0&0.563488 & 750.0 \\ % ld3
  T85r3 &0.855& 536\,870\,912 & 981.0&0.568673& 2\,885.0\\ % ld2
  T85r4 &0.855& 536\,870\,912 & 978.0&0.573923& 3\,250.0\\ % ld
% T85r5 does not really add anything, was too short
%  T85r5 &0.855& 536\,870\,912 & 975.0&0.579237& 1\,500.0\\ % wang
\hline 
  T85r2h &0.855& 67\,108\,864& 492.0&0.563488 &  1\,500.0\\ % ld3hs2
  T85r2q &0.855& 8\,388\,608& 246.0&0.563488 & 4\,162.5\\ % ld3qs2
\hline 
T8r1 &0.80& 536\,870\,912 & 957.0&0.612537& 300.0 \\
T8r2 &0.80& 536\,870\,912 & 954.0&0.618336& 437.5 \\
T8r3 &0.80& 536\,870\,912 & 951.0&0.624207& 2\,375.0\\
 \hline
  T7r1 &0.70& 536\,870\,912 &  921.0& 0.687212&  1375.0\\
  T7r2 &0.70& 536\,870\,912 &  920.5& 0.688333& 1300.0\\
  T7r3 &0.70& 536\,870\,912 &  919.5& 0.690581& 667.5\\
  T7r4 &0.70& 536\,870\,912 &  918.0& 0.693972& 1212.5\\
  \hline
  T6r1 &0.60& 536\,870\,912 &  898.0& 0.741380& 300.0\\
  T6r2 &0.60& 536\,870\,912 &  897.0& 0.743862& 328.0\\
  T6r3 &0.60& 536\,870\,912 &  896.0& 0.746356& 600.0\\
\end{tabular}
\end{ruledtabular}\label{tab:t1}
\end{table}

\subsection{Equilibrium simulations}\label{sec:eq}

Comparing the results of our bubble nucleation simulations with theoretical models requires knowledge of the
thermodynamic properties of the TFS-LJ fluid, especially the vapor and liquid densities at saturation ($\rho\sub{v}$ and $\rho\sub{l}$), 
the pressure at coexistence $P_{\rm eq}$ and the planar surface tension $\gamma$. Some of these values can be found in the literature\cite{Wang2008}, but the statistical uncertainties in $\gamma$ are too large for our purposes.

To estimate these quantities we performed a series of equilibrium MD simulations of liquid-vapor systems. At
each temperature one relatively large system (5 million atoms) consisting of a liquid slab with
vapor on both sides was set up and equilibrated for a large number of time-steps (typically one million)
at fixed temperature (NVT) and with periodic boundaries (see refs. \citep{Baidakov2007,paper2} for details).

When the system has reached a perfectly stable equilibrium state it is evolved for another one million NVT time-steps. At every time-step
the Kirkwood-Buff pressure tensor across the two planar vapor-liquid interfaces is calculated, and the time-averaged
values stored every 50\,000 steps, and converted into estimates of $\rho\sub{v}$, $\rho\sub{l}$, $P_{\rm eq}$ and $\gamma$. This
is done twenty times to determine the average values and the one-sigma scatter for these quantities. The large simulation size and long time integration leads to less statistical uncertainty compared to earlier smaller and shorter simulations. We can then constrain the planar surface tension $\gamma$ within a relative one sigma error of $2.6\%$ or less, while earlier estimates have larger uncertainties of around 10$\%$. However, even the remaining small uncertainties in the surface tension lead to uncertainties of around one order of magnitude in the theoretical nucleation rate predictions (see Figure \ref{Jmd} and Table \ref{tab:t3}).
 
At $T = 0.855 \epsilon/k$ we can compare our values to earlier estimate for the same TFS-LJ fluid at the same temperature: 
Wang et al. \citep{Wang2008} found $\gamma = 0.098 \pm 0.008$, which agrees well with our estimate within their large statistical uncertainty.
The estimate from density functional theory (DFT) $\gamma = 0.119$ [\cite{Lutsko2006}] however, seems too high. Both estimates
are higher than ours and therefore lead to significantly lower model predictions for the nucleation rate. The far better agreement with the classical models we find here is caused
not only to our lower measured rates, but also by our lower surface tension estimates, which increase the theoretical nucleation rate predictions by large factors.

\begin{table}
\caption{Equilibrium properties of the TFS-LJ fluid: Pressure $P_{\rm eq}$, densities of vapor and liquid phase, $\rho\sub{v}$ and $\rho\sub{l}$,
Poynting correction $\delta$ and planar surface tension $\gamma$.
 Values in parenthesis are one sigma errors in the
last given digit(s). These values were obtained from MD simulations of a liquid slab in equilibrium with the vapor phase by
calculating the Kirkwood- Buff pressure tensor across the two planar vapor-liquid interfaces, see refs. \cite{Baidakov2007,paper2} for details.}
\begin{ruledtabular}
\begin{tabular}{ l c c c c c c c c c} $T $ &    
 $P_{\rm eq}$ & 
 $\rho\sub{v}$  &  
 $\rho\sub{l}$  &  
 $\delta$ &
 $\gamma$ & 
\\ $\epsilon/k$& $ [\epsilon/\sigma^3]$&  $[m/\sigma^3]$& $[m/\sigma^3]$& &$ [\epsilon/\sigma^2]$ &
\\  \hline     
0.855 &  0.04610(8)& 0.0833 & 0.595 & 0.870 &0.0895(24)  \\ \hline
 0.8  &0.03028(6)& 0.0505 & 0.652& 0.926 &0.168(3) \\ \hline
 0.7  &0.01186(7)& 0.0198 & 0.729& 0.973&0.329(4) \\ \hline
 0.6  &0.00337(4)& 0.00606 & 0.792& 0.992 &0.511(5) \\
 \end{tabular}
\end{ruledtabular}\label{tab:t2}
\end{table}

\section{Nucleation Rates}\label{sec:rates}

Figures \ref{fig:treshT855} to \ref{fig:treshT6} show the number of bubbles larger than some threshold volume as a function of
time. Using the threshold method, these counts are used to derive nucleation rate estimates as described in Section \ref{Jmeas}.
The steady state bubble nucleation rates obtained from our direct NVE MD simulations are given in Table \ref{tab:t3}, and in the
next two sections we compare them to predictions from classical models and to earlier estimates from smaller simulations.

\subsection{Comparison with classical models}

\begin{figure*}
\includegraphics[height=.75\textheight]{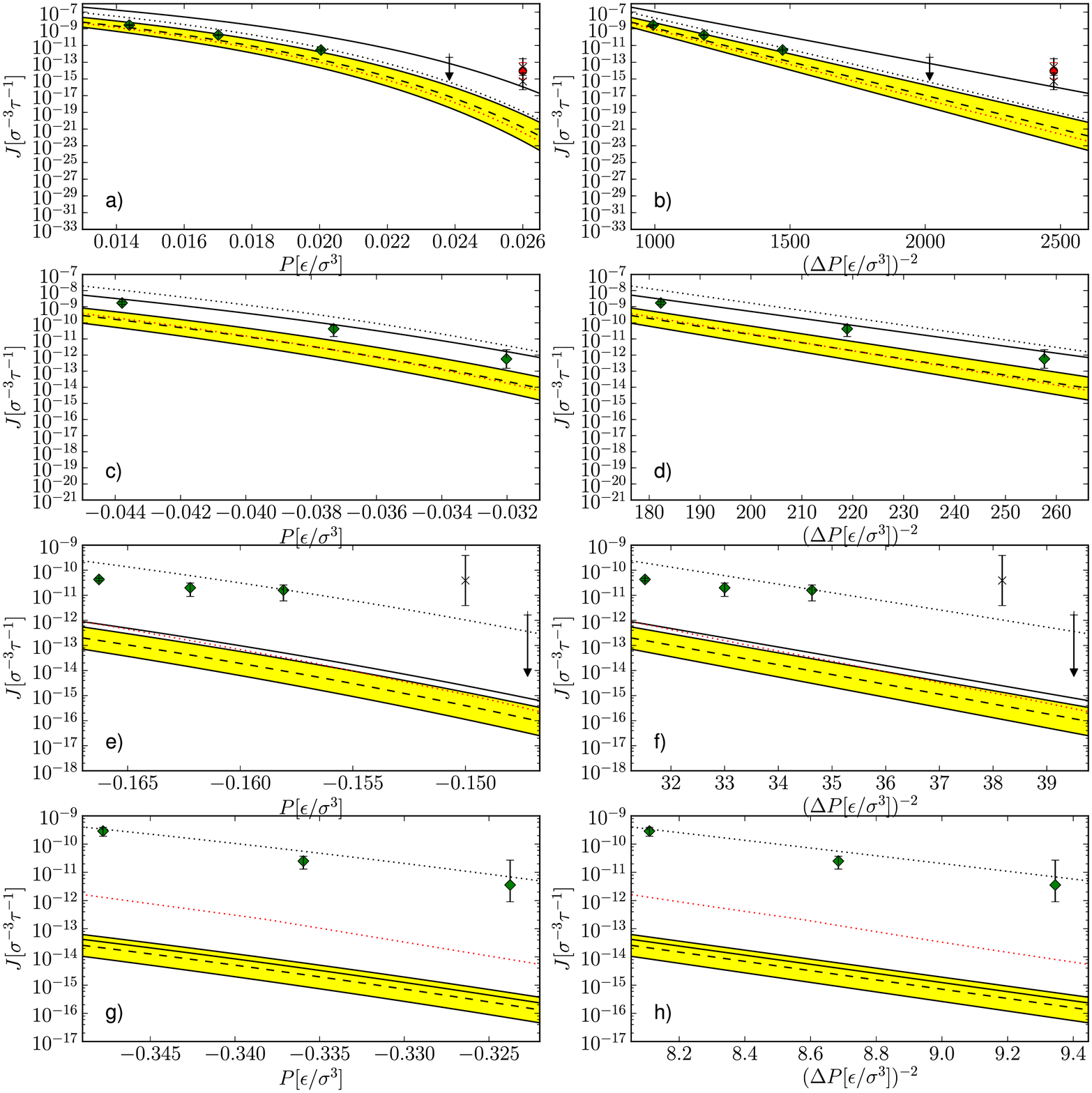}
\caption{(Color online) Measured nucleation rates at $kT/\epsilon$ = 0.855 (panels a and b), 0.8 (c,d), 0.7 (e,f) and 0.6 (g,h) against ambient pressure (left column) and against the inverse of the pressure difference squared (right column). Our MD results (diamonds, down arrows for upper limits) are compared with the CNT (solid lines) and PCNT (dashed lines) predictions and to the MD FFS simulation results from  Wang et al.\cite{Wang2008} (circles) and Meadley $\&$ Escobedo\cite{Meadley2012} (crosses). The shaded areas around the PCNT model predictions illustrate the effect of the one sigma uncertainty in our planar surface tension estimates. The dotted lines are PCNT-like predictions with bubble size dependent
surface tensions: black (upper) dotted lines show the Tolman correction with $\delta_T = 0.25  \sigma$, red (lower) dotted lines use the surface tension from
eq. \ref{tolmanbb}.
}\label{Jmd}
\end{figure*}

It is widely believed that classical models generally underestimate bubble nucleation rates by very large factors\cite{Zeng1991,Wu2003,Wang2008,Sekine2008,Meadley2012} and that they perform significantly worse than in
the case of vapor-to-liquid nucleation\cite{Zeng1991}. However, we do find very good agreement with the classical theory (both with CNT and PCNT) at high temperatures ($T = 0.8 \epsilon/k$ and $0.855 \epsilon/k$).
The classical nucleation rate predictions match quite well in the superheated boiling regime ($T = 0.8 \epsilon/k$), and also for moderate cavitation cases (slightly negative pressures and $0.8 \epsilon/k$), see Figure \ref{Jmd}.

At lower temperatures the measured nucleation rates are significantly higher than the CNT predictions, the ratio $J/J_{CNT}$ reaches values around $10^4$ at $T = 0.6 \epsilon/k$. As in the case of vapor-to-liquid condensation,
CNT seems to predict realistic rates at some intermediate or high temperature, and to underestimate the rates below that temperature\cite{paper1}. And the predicted temperature dependence is off by a similar amount: decreasing
the temperature by  $10\%$ increases the ratio $J/J_{CNT}$ by about one order of magnitude in both the bubble nucleation and the vapor-to-liquid case\cite{paper1} in the nucleation rate regime accessible to large scale MD simulations.

\subsection{Implications for the size dependence of the surface tension}\label{sec:tolman}

The large ratios of measured-to-predicted nucleation rates $J/J_{PCNT}$ at low temperatures may indicate that the surface tension depends on bubble size.
Bubbles near the critical scale determine the nucleation rates, and this scale is smaller at low temperatures. A suitable correction to the planar surface tension\citep{core-surface,Baidakov2014}, which results in a lower
surface tension for small bubbles would lead to a smaller energy barrier and higher nucleation rates at low temperatures.

Assuming that the classical models are correct except for the assumed size independent surface tension, one can use the nucleation rates from the simulations for an indirect estimate
of the surface tension of critical size bubbles\cite{Baidakov2014}. Using the reconstructed free energy landscapes (Figures \ref{deltag855} and \ref{deltag8}) such estimates can in principle
be made for all sizes represented in the simulated size distribution\cite{paper3}.

The correction introduced by Tolman\cite{Tolman1949} is
\begin{equation}\label{tolman}
\gamma_T (r)  = \gamma / (1 + 2 \delta_T/r) \simeq \gamma (1 - 2 \delta_T/ r) \; .
\end{equation}
A small, positive Tolman length of $\delta_T = 0.25 \sigma$ would lead to perfect agreement at low temperatures ($J/J_{PCNT} \simeq 1$ at $T = 0.6 \epsilon/k$ and $0.7 \epsilon/k$), and it would not spoil the good
agreement between the MD simulations and the classical models at the higher temperatures ($J/J_{PCNT} \simeq 0.1$ at $T = 0.8 \epsilon/k$ and $0.855 \epsilon/k$), see Figure \ref{Jmd}.
This agrees well with the Tolman length of $\delta_T = 0.5 r_0 \simeq 0.32  \sigma$ derived from large scale vapor-to-liquid nucleation simulations\cite{paper3},
using the somewhat different standard LJ potential with $r_{\rm cut} = 5 \sigma$. Our estimate of $\delta_T = 0.25  \sigma$ is also consistent with the constraint
$| \delta_T | \le 0.5 \sigma$ obtained by Horsch et al. \cite{Horsch2012} for equilibrium TFS-LJ droplets,
with the practically constant $\gamma$ found for equilibrium LJ-bubbles with $r > 6\sigma$ by Matsumoto and Tanaka\cite{Matsumoto2008},
but not with the stronger size dependence reported in Akbarzadeh et al. \cite{Akbarzadeh2011}. Note however that estimating the surface tension of small equilibrium MD-nanobubbles
is extremely challenging\cite{Cosden2011}. 

By cutting out spherical voids of various sizes in molecular dynamics simulations of a liquid under negative pressure, and observing whether the resultant bubble is stable, \cite{Kuskin2010} estimate a Tolman length $\delta_T = 0.26\pm0.01\,\sigma$ at $kT=0.6$. This is very close to our estimate of $\delta_T = 0.25\epsilon$ at $kT=0.6\epsilon$.

\begin{figure}
\includegraphics[height=.32\textheight]{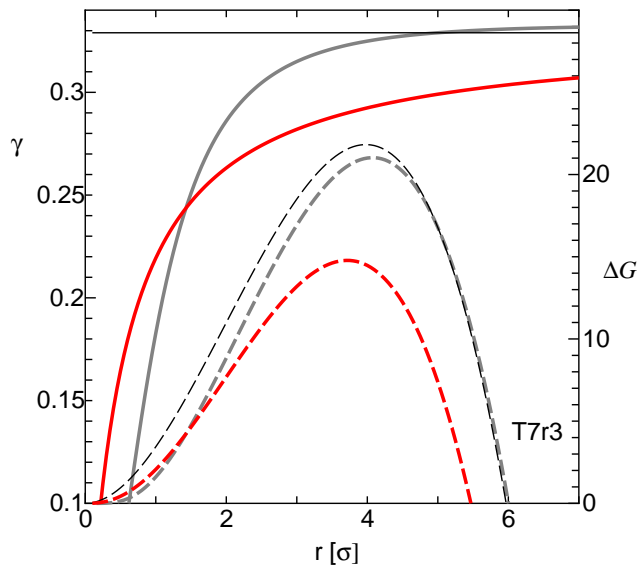}
\caption{(Color online) Surface tensions (solid lines) as a function of
bubble radius at $T = 0.7 \epsilon/k$: The planar surface tension (thin solid), 
a small, positive Tolman length of $\delta_T = 0.25 \sigma$ (red solid line, lower one at large radii) and
using eq. \ref{tolmanbb} with $\delta_B = - 0.1 \sigma$ and $l$=1.0 $\sigma^2$ (gray solid line, upper one at large radii).
Dashed lines show the corresponding free energy curves for run T7r3 (see axis label on the right):
PCNT assumes the constant planar surface tension (thin). The
other two $\Delta G$ curves use a small, positive Tolman length (red) and eq. \ref{tolmanbb} (gray).
}\label{delg-rad}
\end{figure}

Baidakov and Bobrov\cite{Baidakov2014} use a different approximation, which is applicable to a wider range of bubble sizes:
\begin{equation}\label{tolmanbb}
\gamma_T (r)  = \gamma / (1 + 2 \delta_B/r  + l^2 / r^2)  \; ,
\end{equation}
and includes a second parameter $l$. With similar parameters as in\cite{Baidakov2014} ($\delta_B = - 0.1 \sigma$ and $l$=1.0 $\sigma^2$) the predicted nucleation rates at low temperatures are still too low by 
two to three orders of magnitude. The correction is too small at the critical sizes of 3 to 4 $\sigma$, which are relevant for the nucleation rates in our low temperature runs (see Figure \ref{delg-rad}).
The correction becomes large at the smaller critical
scales of 1 to 2 $\sigma$ relevant in the runs in Badiakov and Bobrov\cite{Baidakov2014}, and may be specific to the regime of nucleation rates, temperatures and inter-particle potential probed in their simulations.

These corrections likely depend on temperature. A small, positive Tolman length of $\delta_T = 0.25 \sigma$ fits the nucleation rates well at all temperatures, but it would
lead to worse fits to the free energy landscapes reconstructed from the bubble size distribution at higher temperatures (Figures \ref{deltag855} and \ref{deltag8}).
A size-independent surface tension 
(as found in\cite{Matsumoto2008}) actually matches the results from our high temperature runs better.

\subsection{Comparisons with other simulations}

\begin{figure}
\includegraphics[height=.3\textheight]{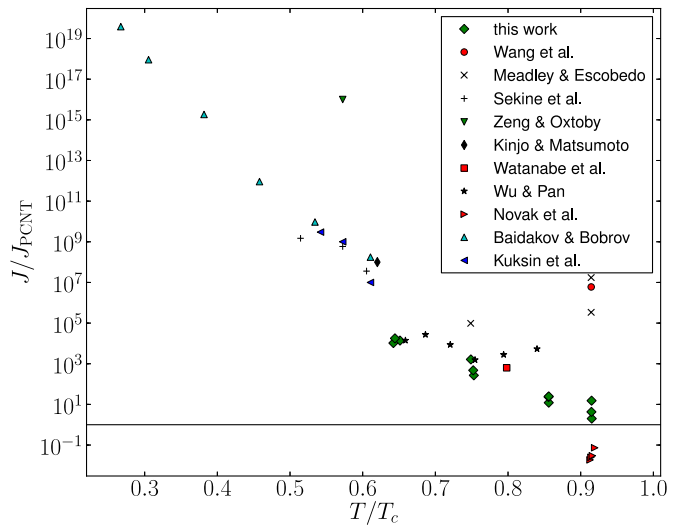}
\caption{(Color online) Measured nucleation rates divided by the PCNT estimate against $T / T_c$ for our
simulations (green diamonds) and bubble nucleation results from the literature.}\label{JmdJpcnt}
\end{figure}

Figure \ref{JmdJpcnt} shows the measured nucleation rates divided by the PCNT estimate as a function of temperature relative to the critical temperature.
The relative temperature scale makes comparisons between results from simulations which use different intermolecular potentials possible: 
Our direct MD results and the MD FFS simulation results from Wang et al.\cite{Wang2008} and Meadley $\&$ Escobedo\cite{Meadley2012}
employed a  TFS-LJ intramolecular potential with a cut-off scale of $r_{\rm cut} =  2.5 \sigma$, which results in a fluid with $T_{c} = 0.935$.
The other studies used standard LJ potential without a force shift and with cut-off scale of from $r_{\rm cut} =  2.5 \sigma$ to $r_{\rm cut} =  6.578 \sigma$,
which gives rise to critical temperatures from $T_{c} = 1.19$ to $T_{c} = 1.31$ \cite{Trokhymchuk1999}.

For the comparison we calculated $J/J_{CNT}$ using the both the $J$ and the $J_{PCNT}$ (or $J_{CNT}$, the differences are very small at low temperature) values given in each paper.
Only for references \cite{Wang2008,Meadley2012},
which simulated the exact same TFS-LJ at the same temperatures as this work, do we recalculate $J_{PCNT}$. Note the uncertainties in the $J_{CNT}$ and $J_{PCNT}$ are quite large, mostly due to the uncertainties in the planar surface tension estimates $\gamma$. Figure \ref{JmdJpcnt} does not show these errors. For the case of our $J_{PCNT}$ estimates, they are illustrated by the shaded regions in Figure \ref{Jmd}.

The majority of bubble nucleation studies find significantly larger nucleation rates than predicted by CNT and PCNT, i.e. $J/J_{CNT}$ of $10^5$ and larger. This seems to be in qualitative agreement with the density functional calculation by Zeng \& Oxtoby \cite{Zeng1991}, which predict a very large $J/J_{CNT} \sim 10^{16}$. Our 
simulations offer an way to cross check these earlier results, because they are quite different from all the previous simulations and in many aspects more accurate and reliable: We use simple and robust direct NVE MD simulations which are more than a factor of 1\,000 times larger than all previous bubble nucleation simulations (except for the spinodal nucleation simulations of \cite{Watanabe2010}, which contain up to $10^7$ atoms). Our runs are the first which are demonstrably not affected by finite size effects. We find very good agreement with PCNT and only a relatively small increase in $J/J_{CNT}$ as we go to lower temperatures. We do not find the large $J/J_{CNT}$ values found in most earlier studies. Our results agree well with the relatively small deviations from classical models found by Wu and Pan\cite{Wu2003}, Watanabe et al. \cite{Watanabe2010} and Novak et al.\cite{Novak2007}.

Some of the differences between the simulation results shown in Fig. \ref{JmdJpcnt} may be due to the different intermolecular potentials used. For example the standard LJ potential with $r_{\rm cut} =  6.578 \sigma$ used in Baidakov $\&$ Bobrov \cite{Baidakov2014} has a far larger planar surface tension. Even after rescaling with the corresponding critical temperature $T_c$, the differences remain large: at 0.76 $T_c$ for example, the $\gamma$ value of the force-shifted $r_{\rm cut} =  2.5 \sigma$ potential we use here is
about a factor of two smaller. A larger surface tension leads to smaller critical bubble sizes, it is therefore plausible that the numerical experiments in
Baidakov $\&$ Bobrov \cite{Baidakov2014} and similar studies probe a different regime, where bubble curvature and atomistic effects play
a larger role and where classical models are therefore less accurate. Their size dependent fit to the bubble surface tension actually leads to good fits to our nucleation
rates (see section \ref{sec:tolman}), their larger $J/J_{CNT}$ values may therefore be due to the smaller critical bubbles in the regime probed in their simulations.

Other studies \cite{Wang2008,Meadley2012} do use the exact same potential at very similar thermodynamic states, and can be compared directly with our results: Both find significantly larger $J/J_{PCNT}$ factors than we do (see also Figure \ref{Jmd}). Meadley $\&$ Escobedo\cite{Meadley2012} show that their results are affected by the size of the simulations, their nucleation rates decrease when 8\,000 instead of 3\,375 atoms are used. $J$ was likely overestimated due to the small simulations sizes used in these studies\cite{Meadley2012} and perhaps also due to artefacts in the FFS method\cite{Torabi2013}.

\section{Critical sizes from the first nucleation theorem}\label{sec:crit_sizes}

The first nucleation theorem can be used to calculate the volume of critical bubbles from the nucleation rates directly \cite{Ford1997,Kashchiev2000, Wilemski2006}. It does not rely on a specific model (like CNT for example), and provides a useful independent estimate of the critical bubble size. Wilemski\cite{Wilemski2006} showed that the critical volume is
\begin{equation}\label{nucltheo}
V_{c} =  \frac{\rho_{\rm L}}{\rho_{\rm L} - \rho_{\rm V}} \left( \frac{\partial \ln J}{\partial P_{\rm L}} \right)_T   \; .
\end{equation}

We estimate the derivative by taking the finite differences to the next available nucleation rate at the same temperature. If both a higher and a lower rate are available, these two rates are used to calculate the slope. Uncertainties in the nucleation rate estimates $J_{\rm MD}$ are propagated into an error estimate on the critical sizes by using the shallowest and steepest slopes allowed by the one sigma errors on $J_{\rm MD}$. The resulting estimates of $V_c$ are converted into a spherical radius $r_c$, and listed in Table \ref{tab:t3}. 

We find that the critical bubble sizes from the nucleation theorem are in very good agreement with the predictions from both
classical models, CNT and PCNT at all temperatures. While the predicted nucleation rates agree with the simulations only
at the higher temperatures, the pressure dependence (i.e. the slopes in the left hand panels of Figures \ref{Jmd})
of the nucleation rates seems to be predicted accurately by the classical models at all temperatures. Similar
conclusions have been reached in the case of vapor-to-liquid nucleation in refs. \cite{Wedekind2007,Tanaka2005,Tanaka2011,paper1}.

\begin{table*}\caption{Average liquid temperature $T$ and pressure $P$ during the NVE integration (up to the final pressure increase),
pressure difference $\Delta P$ (equilibrium pressure $P_{\rm eq}$
minus measured pressure $P$), critical bubble radius $r^*$ and nucleation rate $J$ for each run.
Our critical sizes $r_{\rm{c}}^*$ were derived form the measured rates $J_{\rm{MD}}$ using the nucleation theorem, Eq. (\ref{nucltheo}).
The $r_{\rm{c}}^*$ estimates from W08\cite{Wang2008} and ME12\cite{Meadley2012} are based on the FFS method.
The nucleation rates were derived using CNT and PCNT and measured directly in the MD simulations. The error bars on the model predictions
come from the uncertainties in the surface tensions (see Table II).}
\begin{ruledtabular}
\begin{tabular}{l |ccc|ccc|ccc}
  Run ID & $T$ & $P$ &  $ \Delta P $ &
$r_{\rm{c}}^*$& $r_{\rm{CNT}}^*$&$r_{\rm{PCNT}}^*$& 
$J_{\rm{MD}}$ &
$J_{\rm{CNT}}$ &
$J_{\rm{PCNT}}$ \\
 & $[\epsilon/k]$& $[\epsilon/\sigma^3]$ & $[\epsilon/\sigma^3]$  &$[\sigma]$ &$[\sigma]$& $[\sigma]$& $ [ \sigma^{-3}\tau^{-1} ]$& $ [ \sigma^{-3}\tau^{-1} ]$
 &$ [ \sigma^{-3}\tau^{-1} ]$\\ 
\hline 
T85r1 &0.8557&0.01438 & 0.03172 & 6.3$\pm$0.2 & 5.64 & 6.49  & 2.8$\pm$0.3$\times10^{-9}$&1.2$\times10^{-7\pm0.5}$&1.4$\times10^{-9\pm0.6}$\\ %ld4: within 11% from 40k to 97k; i_thres=10 gives this higher rate
T85r2 &0.8553&0.01701 & 0.02909 & 6.6$\pm$0.1& 6.15& 7.07 & 1.8$\pm$0.2$\times10^{-10}$ &8.8$\times10^{-9\pm0.6}$ &4.2$\times10^{-11\pm0.8}$\\ % ld3:  within 3.5% from 100k to 204k
T85r3 &0.8555&0.02004 & 0.02606 & 6.9$\pm$0.2& 6.87& 7.90 & 2.9$\pm$0.6$\times10^{-12}$ & 1.5$\times10^{-10\pm0.7}$ &1.9$\times10^{-13\pm1.0}$\\ % ld2: within 2.5% from 42k to 880k
T85r4 &0.8551&0.02383 & 0.02227 &$>$5.1& 8.04 & 9.24  & $<$3.8$\times10^{-13}$ & 7.1$\times10^{-14\pm1.0}$ & 7.8$\times10^{-18\pm1.3}$\\ % ld 
% T85r5 does not really add anything, was too short
% T85r5 &0.8550&0.02839 & 0.01771 & $>$3.4 & 10.1 & 11.6  & $<$8.2$\times10^{-13}$ & 5.0$\times10^{-21\pm1.6}$ & 2.8$\times10^{-27\pm2.1}$\\ % wang
\hline
 W08  &0.8550&0.02600  & 0.0201 & 6.1 & 8.91 & 10.2  & 0.9$\times10^{-14\pm1}$&1.1$\times10^{-16\pm1.2}$&1.5$\times10^{-21\pm1.6}$\\   % Wang et al. 2009, FFS estimates
 ME12a&0.8550&0.02600  & 0.0201 & 6.4(3) & 8.91 & 10.2  & 2.6$\times10^{-14\pm1}$&1.1$\times10^{-16\pm1.2}$&1.5$\times10^{-21\pm1.6}$\\   %Stacey&Escobedo2012, FFS estimates with N=3375
 ME12b&0.8550&0.02600  & 0.0201 & 7.2(5) & 8.91 & 10.2  & 5.1$\times10^{-16\pm1}$&1.1$\times10^{-16\pm1.2}$&1.5$\times10^{-21\pm1.6}$\\   % Stacey&Escobedo2012, FFS estimates with N=8000
\hline
\hline 
T8r1 &0.8004&-0.04379 & 0.07407 & 4.9$\pm$0.4& 4.54 & 4.90  & 1.7$\pm$0.3$\times10^{-9}$&2.9$\times10^{-9\pm0.4}$&1.4$\times10^{-10\pm0.5}$\\ %n6125: 40k to 92k; 2.5%
T8r2 &0.8001&-0.03713 & 0.06741 & 5.2$\pm$0.3& 4.98 & 5.39  & 4.2$\pm$2.8$\times10^{-11}$&6.9$\times10^{-11\pm0.5}$&1.8$\times10^{-12\pm0.6}$\\ %n618: 40k to 145k; 1.5%
T8r3 &0.8004&-0.03202 & 0.06230 & 5.2$\pm$1.0 & 5.39 & 5.83  & 1.5 - 22$\times10^{-13}$&1.6$\times10^{-12\pm0.6}$&2.3$\times10^{-14\pm0.7}$\\ %n624: 50k (took a bit longer!) to 760k; 1.0%
\hline
\hline 
T7r1 &0.7040&-0.16627 & 0.17813 &3.1$\pm$1.0& 3.69 & 3.80  & 4.3$\pm$0.5$\times10^{-11}$&7.2$\times10^{-13\pm0.4}$&1.6$\times10^{-13\pm0.4}$\\ %n69: 22k to 56k; 4%
T7r2 &0.7032&-0.16222 & 0.17408 &3.0$\pm$1.0& 3.78 & 3.88  & 2.0$\pm$1.1$\times10^{-11}$&2.0$\times10^{-13\pm0.4}$&4.2$\times10^{-14\pm0.5}$\\ %n692: 26k to 115k; 10%
T7r3 &0.7002&-0.15808 & 0.16996 &2.8$\pm$0.8& 3.87 & 3.98  &1.6$\pm$1.0$\times10^{-11}$& 5.1$\times10^{-14\pm0.5}$& 9.8$\times10^{-15\pm0.5}$\\ %n695: 42k to172k;  1.5%
T7r4 &0.7000&-0.14724 & 0.15910 & $>$3.0& 4.14 & 4.25  & $<$ 1.6$\times10^{-12}$&7.9$\times10^{-16\pm0.5}$&1.2$\times10^{-16\pm0.6}$\\ %n70: 42k to 350k; no nucleation
\hline
ME12&0.7000&-0.15000 & 0.16186 & 3.1 & 4.06 & 4.18  & 3.9 $\times10^{-11\pm1}$&2.5$\times10^{-15\pm0.5}$&4.0$\times10^{-16\pm0.5}$\\ % FFS estimates
\hline 
T6r1 &0.6090&-0.34779 & 0.35116 &3.1$\pm$0.4& 2.91 & 2.93  &2.9$\pm$1.0$\times10^{-10}$&3.4$\times10^{-14\pm0.4}$&2.1$\times10^{-14\pm0.4}$\\ %l898: 20k to 42k; 1.5%, ~15 bubbles
T6r2 &0.6005&-0.33596 & 0.33933 &2.9$\pm$0.4& 3.01 & 3.03  &2.5$\pm$1.2$\times10^{-11}$&4.0$\times10^{-15\pm0.4}$&2.4$\times10^{-15\pm0.4}$\\ %l897: 20k to 68k; 1.5%; 3 bubbles
T6r3 &0.6025&-0.32375 & 0.32712 &2.9$^{+0.6}_{-2.9}$& 3.12 & 3.15  &0.9 - 14$\times10^{-12}$&3.4$\times10^{-16\pm0.4}$& 2.0$\times10^{-16\pm0.4}$\\ %l896: 20k to106k;  1.5% ; one bubble
\end{tabular}
\end{ruledtabular}\label{tab:t3}

\end{table*}

\section{Free energy for bubble formation}\label{sec:size_dist}

If the free energy as a function of bubble size $\Delta G(r)$ is known, one can calculate the {\it equilibrium} bubble size distribution directly with Eq. \ref{sizedis}. During steady state nucleation, this distribution is only realized for small bubble sizes ($r \ll r_c$). The {\it steady-state} distribution turns out to be a factor of two lower at $r_c$ and flat for larger bubble sizes\cite{Kelton2010}. Conversely, one can measure the steady-state size distribution and the nucleation rate in a large scale MD simulation and use them to reconstruct the entire free energy function $\Delta G(r)$, even at $r_c$ and above\cite{paper3}.

\begin{figure}
\includegraphics[height=.55\textheight]{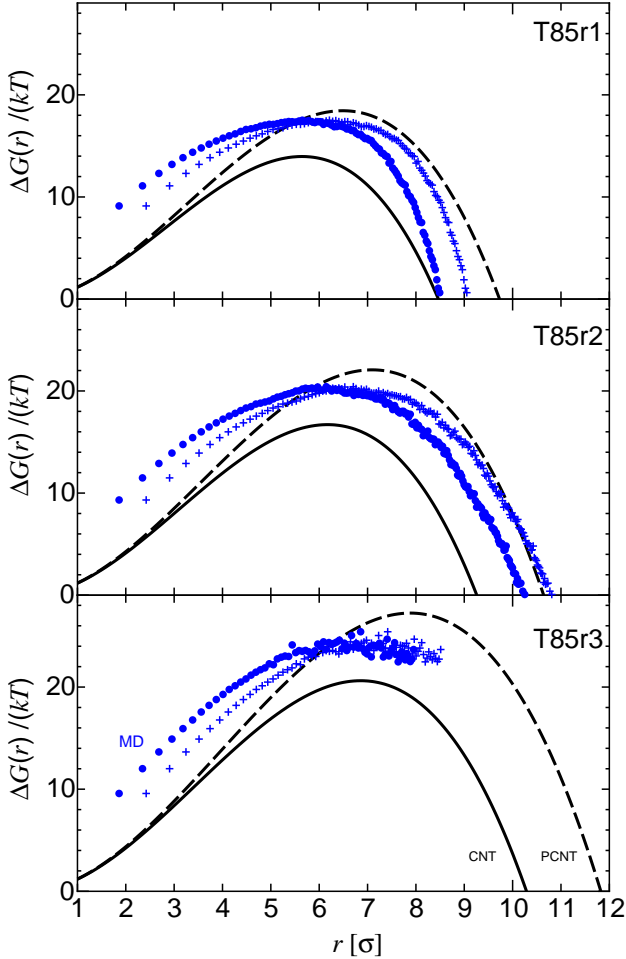}
\caption{(Color online) Free energy for bubble formation at $kT/\epsilon$ = 0.855. Reconstructed free energy curve from the MD bubble size distribution 
in terms of $r_{\rm MD}$ (circles) and in terms of the estimated equimolar radius ($r_{\rm MD}+0.56\sigma$) (crosses). Model predictions
are shown with solid (CNT) and dashed (PCNT) lines.
}\label{deltag855}
\end{figure}

\begin{figure}
\includegraphics[height=.38\textheight]{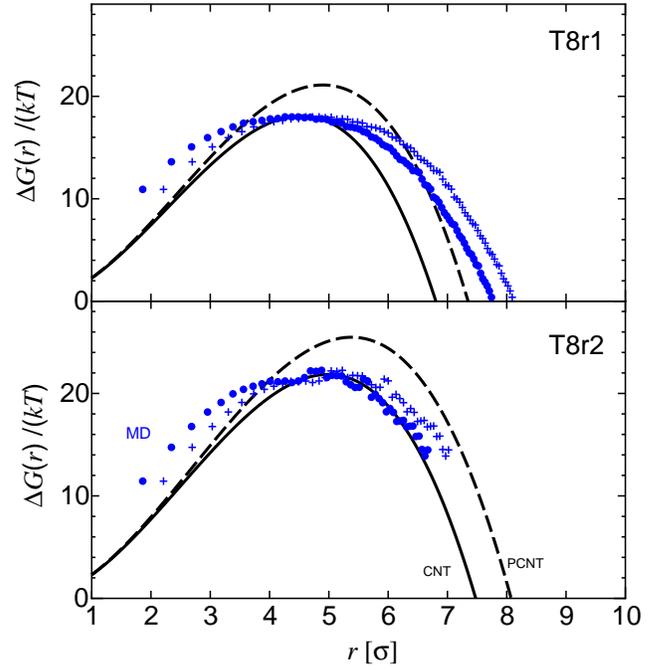}
\caption{(Color online) Free energy for bubble formation at $kT/\epsilon$ = 0.8. Reconstructed free energy curve from the MD bubble size distribution 
in terms of $r_{\rm MD}$ (circles) and in terms of the estimated equimolar radius ($r_{\rm MD}+0.35\sigma$) (crosses). Model predictions
are shown with solid (CNT) and dashed (PCNT) lines.}\label{deltag8}
\end{figure}

Figures \ref{deltag855} and \ref{deltag8} show the free energy for bubble formation. The bubble size distributions were measured in the MD simulations and time-averaged over the steady state nucleation phase at constant liquid pressure. Together with our MD nucleation rate measurement (and assuming that the empirical pre-exponential factor in Eqn. (\ref{J}) is correct) this allows the reconstruction of the full free energy function with the method from Tanaka et al. \cite{paper3}. The bubble sizes were estimated by converting the volume of all cells below the threshold density of 0.2 $m/\sigma^3$ into a spherical radius $r_{\rm MD}$. Comparing these radii with the equimolar radii of a wide range of large, stable bubble reveals that our $r_{\rm MD}$ values are slightly smaller, by an offset which depends on temperature, but not on bubble size. In figures \ref{deltag855} and \ref{deltag8} we use both $r_{\rm MD}$ as well as the estimated equimolar radii.

The reconstructed free energy landscape agrees well with CNT and PCNT at these high temperatures. At the lower temperature the bubbles are too small to allow a detailed comparison with this method. The peak position ($r_c$) agrees with PCNT at both temperatures. The peak height matches the PCNT prediction perfectly at $kT/\epsilon$ = 0.855, and is slightly lower than PCNT at $kT/\epsilon$ = 0.8. This is consistent with the excellent agreement of the nucleation rates at $kT/\epsilon$ = 0.855, and the slightly larger $J_{\rm MD}$ versus $J_{\rm PCNT}$ at $kT/\epsilon$ = 0.8 (Figure \ref{Jmd}).

The models assume mechanical equilibrium at all sizes, as well as a sharp density transition from vapor to liquid, while our simulated bubbles have transition regions which are several $\sigma$ wide\cite{Angelil2014}, and their exact radii are difficult to define. For comparisons with classical models one often uses the equimolar radius.  Indeed, the curves shifted to larger radii to approximate the equimolar radii match the predictions slightly better. The agreement is surprisingly good, given the many differences between the model assumptions and the actual properties of our MD bubbles (wide interfaces, non-spherical shapes, non-isothermal effects, etc. \cite{Angelil2014}).

\section{Local hot spots}\label{sec:local_hot_spots}

In a recent study Wang et al. \cite{Wang2008} reported that in their MD simulations the occurrence of local temperature fluctuations (``''hot spots'') correlates strongly with subsequent bubble formation. This process is not present in the CNT, and may explain the far larger nucleation rates found in their MD simulations in comparison to the CNT prediction. However, at the same temperature as used in Wang et al. (T=0.855$\epsilon/k$) we find far lower rates and relatively good agreement with CNT (see Section \ref{sec:rates}). In this section we show that there are no local hot spots preceding bubble nucleation in our simulations. There only is a small amount of extra kinetic energy in bubble forming locations, which is perfectly consistent with the amount of movement required to make room for the bubbles within the liquid.

\begin{figure}
\includegraphics[height=.28\textheight]{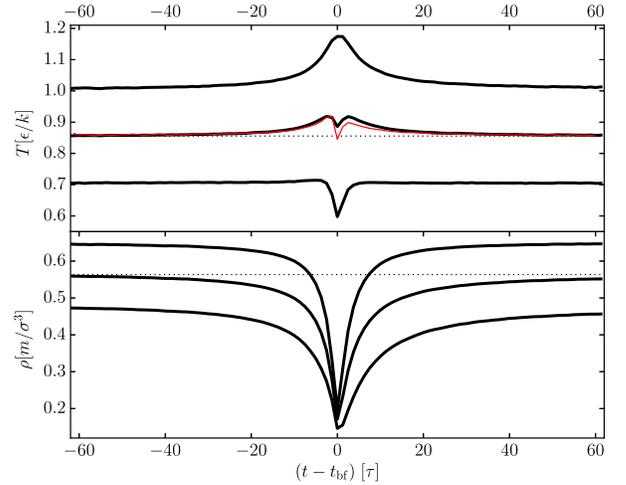}
\caption{(Color online) Average local temperature (top panel) and density (bottom panel) in bubble forming cells of size $(3 \sigma)^3$ as a function of time since the moment of bubble formation $t_{\rm bf}$ (thick solid lines). The upper and lower thick solid lines indicate the one sigma cell-to-cell scatter, dotted lines the averages over the entire simulation volume. The estimated effect of extra kinetic energy from growing and shrinking bubbles is shown with a red thin solid line, see text for details. 
}\label{avgt3s}
\end{figure}

Local kinetic energies and densities were measured for 200\,000 time-steps during the steady-state nucleation regime in run T85r2hs. Both quantities were measured in cells of size $(3 \sigma)^3$ at every step, time-averaged over 500 steps and then stored on disk. Cells with a density below 0.2 are identified as bubble forming cells, and the first time the density falls below the threshold is stored as the moment of bubble formation $t_{bf}$. In this way 234\,000 bubble-forming cells were found, corresponding to $5.3\%$ of the 4.4 million cells which cover the entire simulation volume. For these cells we averaged the local kinetic energies and densities at the same times relative to their moment of bubble formation $t_{bf}$ and plot these averages as a function of $(t-t_{bf})$, see Figure \ref{avgt3s}. 

Near the moment of bubble formation $t_{\rm bf}$, the average kinetic energy in bubble forming cells is indeed slightly higher than in the entire simulation volume. The excess becomes as large as $7\%$, and is comparable to the hot spot signal reported in Fig. 6 in Wang et al. \cite{Wang2008}. However the excess kinetic energy does not precede bubble formation: it appears at the same time as the averaged density drops, and it exactly matches the amount of liquid movement required to obtain these lower densities.

\begin{figure}
\includegraphics[height=.28\textheight]{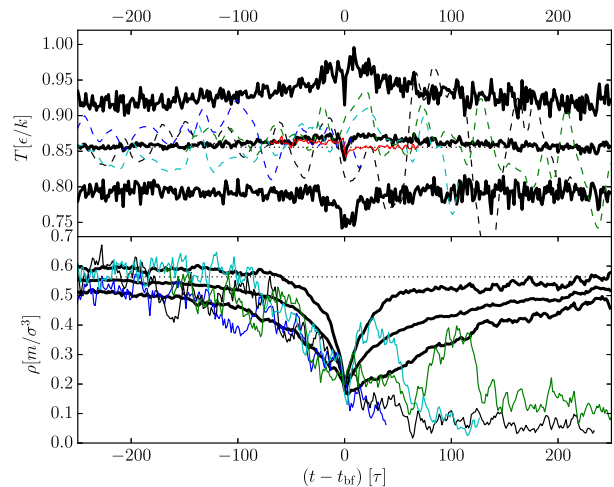}
\caption{(Color online) Same as Fig. \ref{avgt3s} but using larger cells of size $(6 \sigma)^3$. The values in the central cells of four bubble nucleation events are given with dotted lines (top panel) and thin solid lines (bottom panel), see text for details.
}\label{avgt}
\end{figure}

A simple model can match and explain this excess kinetic energy: assuming a spherical vapor bubble with a sharp interface, fully enclosed within the cubic cell. The vapor is assumed to be at the equilibrium density ($\rho_v = 0.0833 m/\sigma^3$) and at a slightly lower temperature (T=0.80$\epsilon/k$), both in agreement with the vapor properties we measure in large bubbles\cite{Angelil2014}. The liquid is assumed to have the same average density and temperature as the entire simulation. At each time $(t-t_{bf})$ a bubble radius is calculated, so that the total density in the cell matches the average density measured in the simulation. Mass conservation implies by how much each radial shell of the liquid around the bubble has to move as the bubble radius changes with time. These velocities are added to the thermal motion in the radial direction, the total kinetic energy of the vapor and liquid in the cell is calculated and
then converted into an average cell temperature (thin lines in the top panels of Figures \ref{avgt3s} and \ref{avgt}).

Despite several unrealistic assumptions (sharp interface, spherical shape, fully enclosed within cell), this simple model reproduces the observed excess kinetic energy very well. We conclude that bubble formation is not preceded by local hot spots. Bubbles form out of an isothermal liquid, as assumed in the CNT. The apparently higher local temperatures in bubble forming cells near $t_{bf}$ are simply caused by the inevitable rearrangement of liquid as bubbles form (and disappear), which introduces some extra kinetic energy into the liquid phase around the bubbles.

Figure \ref{avgt3s} shows the same analysis using larger cells of size $(6 \sigma)^3$. Due to smaller number of cells, the statistics are worse and this figure is noisier than Figure \ref{avgt}.
However, the results are the very similar: The kinetic energy in bubble forming cells is above the simulation average, but well matched by the extra kinetic energy required to grow (or shrink) the cavities required to fit the average local density decrease (or increase). Here only 306 out of 551'368 cells ever fall below the density threshold. A significant fraction of those belong to long-lived or even large, stable bubbles. Therefore the average density in these cells increases quite slowly after $t_{\rm bf}$ and remains below the simulation average even long after $t_{\rm bf}$.

At $t \simeq t_{\rm bf} \pm 2 \tau$ the average local temperatures lie below the simulation average. In large stable bubbles we measure an average vapor density of about T=0.80$\epsilon/k$, below the simulation average of T=0.855$\epsilon/k$ due to the latent heat used during evaporation, see Ang\'elil et al.\cite{Angelil2014}. Simply assuming that these cells contain a mixture of cooler vapor at T=0.80$\epsilon/k$ and isothermal liquid at  T=0.855$\epsilon/k$, both at their equilibrium densities, matches the observed drop in the average local temperatures quite well.

Figures \ref{avgt3s} and \ref{avgt} also indicate that the one sigma cell-to-cell scatter in the local densities and local kinetic energies is quite large. These are real fluctuations, not just sampling noise, since for example the local density in a $(3 \sigma)^3$ cell averaged over 500 measurements has only about  $1\%$ sampling error. Large local temperature fluctuations are present in the entire simulation volume, but they do not correlate at all with subsequent local bubble nucleation events. The excess in the average kinetic energy right before (and also right after) $t_{\rm bf}$ is significantly smaller than a one sigma fluctuation, and it is not a temperature fluctuation (``local hot spot''), but simply caused by the movement required to achieve the density changes.

Four large stable bubbles form during the steady-state nucleation regime in run T85r2h. Their centers move less than 2.5 $\sigma$ during the analysis period and each center lies within one fixed $(6 \sigma)^3$ cell at practically all times. The local densities and kinetic energies of the central cells of these four bubbles (time-averaged over 500 measurements) are shown in Fig. \ref{avgt}: They show wild fluctuations, with an amplitude similar the one sigma scatter found in the other cells. The kinetic energies were smoothed using a Gaussian window with a width of $5\tau$ to make the Figure more easily viewable. The histories of these four large bubbles are obviously very noisy, and no time before $t_{\rm bf}$ can one identify a period were their formation is preceded by a local hot spot. The average temperature of these four regions of successful bubble nucleation fluctuate randomly around the simulation average at all times before the moment of bubble formation $t_{\rm bf}$.

\section{Conclusions}\label{sec:summary}

The main results presented in this work can be summarized as follows:

\begin{itemize}
\item
Large-scale, micro-canonical (NVE) MD runs allow the simulation of homogeneous bubble nucleation in a realistic way,
free of unphysical numerical manipulations (thermostating and barostating) and unaffected by finite simulation size effects.
\item
Bubble nucleation rates are lower than most of the previous estimates from smaller simulations.
\item
Our measured bubble nucleation rates agree well with classical model in the regimes of superheated boiling (positive ambient pressure) and moderate cavitation
(moderately negative pressures): They lie within two orders of magnitude of CNT predictions. The reconstructed free energy landscapes
also agree very well with CNT in these regimes.
\item
In the extreme cavitation regime (at very large negative pressures) CNT does underestimate the nucleation rates significantly.
\item
Introducing a small, positive Tolman length ($\delta_T = 0.25 \sigma$) leads to very good agreement between the predicted and measured nucleation rates at all temperatures. The same conclusion was reached from our recent large scale vapor-to-liquid nucleation simulations\cite{paper3}.
\item
The critical bubbles sizes derived with the nucleation theorem agree well with the CNT predictions at all temperatures. 
\item
Local hot spots reported in an earlier MD simulation\cite{Wang2008} are not seen:
Regions where bubble nucleation events will occur are not above the average temperature, and we observe no correlation between temperature fluctuations and subsequent bubble formation.
\end{itemize}

Our direct large scale bubble nucleation simulations form a large number of stable bubbles in a realistic manner and environment. They produce a lot of additional information about these bubbles, like their shapes,
density and temperature profiles and growth rates. These properties will be presented in a subsequent work\cite{Angelil2014}.

\section{Acknowledgments}
The authors thank V. Kalikmanov for useful discussion. Computations were preformed on SuperMUC at LRZ, on Rosa at CSCS and on the zBox4 at UZH. J.D. and R.A. are supported by the Swiss National Science Foundation.   

\bibliography{paper} 

%\appendix

%\section{viscosity}
%\label{sec:vist}

%Literature summary:

%\cite{rowley1997diffusion}  found $\eta$ values around  0.8 at $\rho = 0.6$, around 1.2 at $\rho = 0.7$ and around 2.0 at $\rho = 0.8$. Viscosity depends only weakly on temperature.

%@article{meier2004transport} found $\eta = 3.0$ to 3.3 for cutoffs from 2.5 to 5.5 at $T= 0.722$and $\rho = 0.8442 $ (near the triple point). Agrees well with many earlier viscosity estimates they cite.

%\cite{galliero2005molecular} found $\eta = 1.778$ and $P=1.042$ at $T= 0.8 T_{\rm crit} = 0.8*1.299 = 1.04$ and $\rho = 2.5 \rho_{\rm crit} = 2.5*0.316 = 0.79$. They use a cuttoff of 2.5 $\sigma$,
%but they use the $T_c$ and $rho_c$ values from another paper based on full LJ potentials. Different potential (12 goes up to 22 and exp instead of LJ) give similar values, mostly a bit larger, up to 2.8.

%\cite{muller1999reversing} found $\eta = 3.2$ to 3.3 with a cutoff at $3.0 \sigma$, at $T= 0.722$ and $\rho = 0.849$ (near the triple point). 

%\end{thebibliography}

\end{document}